\documentclass[aip,apl,reprint,superscriptaddress,preprintnumbers]{revtex4-2}
\usepackage[T1]{fontenc}
\usepackage{times}
\usepackage{graphicx,color}
\usepackage{amsfonts,amsmath,amssymb,amsbsy}
\usepackage[colorlinks=true, linkcolor=blue, citecolor=blue, urlcolor=blue]{hyperref}

\let\mathbf=\boldsymbol

\def\blue#1{\textcolor{blue}{#1}}
\def\magenta#1{\textcolor{magenta}{#1}}
\def\magenta#1{\textcolor{black}{#1}}
\def\emph#1{\textcolor{blue}{#1}}
\def\emph#1{\textcolor{black}{#1}}
\begin{document}

\title{Transformation of a cellular skyrmion to polyomino-like structures}

\author{Jing Xia}
\thanks{Jing Xia and Xichao Zhang contributed equally to this work.}
\affiliation{College of Physics and Electronic Engineering, Sichuan Normal University, Chengdu 610068, China}

\author{Xichao Zhang}
\thanks{Jing Xia and Xichao Zhang contributed equally to this work.}
\affiliation{Department of Applied Physics, Waseda University, Okubo, Shinjuku-ku, Tokyo 169-8555, Japan}

\author{Yan Zhou}
\affiliation{School of Science and Engineering, The Chinese University of Hong Kong, Shenzhen, Guangdong 518172, China}

\author{Xiaoxi Liu}
\affiliation{Department of Electrical and Computer Engineering, Shinshu University, 4-17-1 Wakasato, Nagano 380-8553, Japan}

\author{Guoping Zhao}
\email[Authors to whom correspondence should be addressed:~]{Guoping Zhao, zhaogp@uestc.edu.cn; Masahito Mochizuki, masa_mochizuki@waseda.jp}
\affiliation{College of Physics and Electronic Engineering, Sichuan Normal University, Chengdu 610068, China}

\author{Masahito Mochizuki}
\email[Authors to whom correspondence should be addressed:~]{Guoping Zhao, zhaogp@uestc.edu.cn; Masahito Mochizuki, masa_mochizuki@waseda.jp}
\affiliation{Department of Applied Physics, Waseda University, Okubo, Shinjuku-ku, Tokyo 169-8555, Japan}

\begin{abstract}
Topological spin structures with transformable shapes may have potential implications on data storage and computation. Here, we demonstrate that a square cellular skyrmion on an artificial grid pinning pattern can be manipulated by programmed current pulses. We find that parallel short pulses could result in the elongation of the skyrmion mainly in the current direction, while parallel long pulses are able to induce the elongation in the direction perpendicular to the current due to the intrinsic skyrmion Hall effect. Consequently, a programmed sequence of parallel pulses could lead to the transformation of the skyrmion to I-, L-, and Z-shaped polyomino-like structures without affecting the topological charge. In addition, we find that orthogonal pulses could lead to the transformation to more complex polyomino-like structures, including the T-shaped and irregular ones. Particularly, when a small T-shaped structure is formed, the topological charge of the system is found to be non-integer due to incomplete compensation of local topological charge densities; however, the T-shaped structure is stable on the attractive pinning pattern. Our results offer an effective way to create polyomino-like spin structures toward functional applications.
\end{abstract}

\date{July 10, 2024}

\preprint{\hyperlink{https://doi.org/10.1063/5.0215267}{DOI: 10.1063/5.0215267}}


\maketitle


Skyrmions are topological excitations that can exist in magnetic materials as stable and rigid objects~\cite{Roszler_NATURE2006,Nagaosa_NN2013,Mochizuki_JPCM2015,Wanjun_PR2017,Everschor_JAP2018,Bogdanov_NPR2020,Gobel_PR2020,Reichhardt_RMP2022}.
They have been extensively studied since the first experimental identification of skyrmion lattices in chiral magnets with antisymmetric Dzyaloshinskii-Moriya (DM) interactions~\cite{Muhlbauer_SCIENCE2009,Yu_NATURE2010}.
Isolated skyrmions were created and imagined experimentally later in both multilayers and bulk magnets~\cite{Romming_SCIENCE2013,Wanjun_SCIENCE2015,Woo_NM2016,ML_NN2016,Boulle_NN2016,Soumyanarayanan_NM2017,Wanjun_NPHYS2017,Litzius_NPHYS2017} and were found to be useful for applied spintronic applications~\cite{Finocchio_JPD2016,Kang_PIEEE2016,Fert_NRM2017,Zhang_JPCM2020,Li_MH2021,Luo_APL2021,Marrows_APL2021}, such as the racetrack-type memory~\cite{Sampaio_NN2013,Tomasello_SREP2014,Koshibae_JJAP2015,Yu_NL2017}, multi-state memory~\cite{Wang_NC2020}, and logic computing gate~\cite{Xichao_SREP2015B,Luo_NL2018,Chauwin_PRA2019,Walker_APL2021}.
The motivation of using skyrmions as spintronic components is mainly due to the skyrmionic features such as the nanoscale size, non-volatility, and radiation hardness~\cite{Nagaosa_NN2013,Mochizuki_JPCM2015,Wanjun_PR2017,Everschor_JAP2018,Bogdanov_NPR2020,Gobel_PR2020,Reichhardt_RMP2022,Finocchio_JPD2016,Kang_PIEEE2016,Fert_NRM2017,Zhang_JPCM2020,Li_MH2021,Luo_APL2021,Marrows_APL2021}, which guarantee a high storage density and reasonable retention for information processing.
In recent years, many reports have further suggested the use of skyrmions in non-conventional applications~\cite{Li_MH2021}, which include but not limited to neuromorphic devices~\cite{Huang_N2017,Li_N2017,Pinna_PRA2018,Prychynenko_PRA2018,Jiang_APL2019,Song_NE2020,Lee_PRA2022} and quantum computation~\cite{Psaroudaki_PRL2021,Xia_PRL2022,Psaroudaki_APL2023}.
All these progresses indicate that skyrmions are promising candidates that may revolutionize current hardware information technologies.

In principle, the information processing based on skyrmions could be realized through the manipulation of the degrees of freedom of skyrmions, such as the topological charge, vorticity, and helicity.
For example, the helicity of a frustrated skyrmion could have two degenerate states~\cite{Leonov_NCOMMS2015,Lin_PRB2016A,Leonov_NCOMMS2017,Xichao_NCOMMS2017}, that is, the N{\'e}el-type and Bloch-type states.
The two states can be harnessed for encoding binary information~\cite{Leonov_NCOMMS2015,Lin_PRB2016A,Leonov_NCOMMS2017,Xichao_NCOMMS2017}.
However, in most usual cases the size and shape of skyrmions are invariable properties and are fixed during basic information processing operations~\cite{Reichhardt_RMP2022,Finocchio_JPD2016,Kang_PIEEE2016,Fert_NRM2017,Zhang_JPCM2020,Li_MH2021,Luo_APL2021,Marrows_APL2021}.
Unstable deformation of skyrmions induced by external stimuli may result in the annihilation of skyrmions~\cite{Chen_PRA2022}.

Most recently, it has been suggested that the size and shape of skyrmions could be effectively controlled and manipulated in thin-film and multi-layered systems with artificial pinning landscapes~\cite{Zhang_CP2021,Reichhardt_RMP2022}.
For example, skyrmions on a nanoscale grid pinning pattern may show a square shape, where both the position, size, and shape of the skyrmion can be adjusted in a quantized fashion by current pulses~\cite{Zhang_CP2021}.
The dynamics of skyrmions on pinning patterns are found to depend on both the driving forces and pinning patterns~\cite{Zhang_CP2021,Reichhardt_RMP2022,Reichhardt_PRL2015,Reichhardt_NJP2015,Ma_PRB2016,Muller_NJP2017,Fernandes_NC2018,Menezes_PRB2019,Navau_PRB2019,Navau_NANOSCALE2019,Bhatti_2019,Feilhauer_PRB2020,Lounis_JPCM2020,Lounis_SR2020,Chen_PRA2020,Ishikawa_APL2021,Vizarim_JPCM2021,Souza_PRB2021A,Navau_APLM2022,Souza_NJP2022A,Zhang_PRB2022A,Zhang_NL2023,Souza_JPCM2024A,Souza_PRB2024A,Zhao_SB2024}.
\magenta{Therefore, the interactions between topological spin structures and artificial pinning patterns may unlock more degrees of freedom that can be controlled by external stimuli and thus, may provide new opportunities to facilitate advanced spintronic applications, such as multi-state information storage, computing, and bio-inspired devices.}

\emph{In this work, we show the possibility to realize complex transformation of a cellular square skyrmion to polyomino-like structures in a two-dimensional magnetic layer with an artificial grid pinning pattern.}
\emph{It is worth mentioning that a variety of stable topological multi-solitons taking the form of polyiamonds (i.e., triangular polyominoes) can also be found in the three-dimensional baby Skyrme model~\cite{Sutcliffe_2012}.}

\begin{figure*}[t]
\centerline{\includegraphics[width=0.99\textwidth]{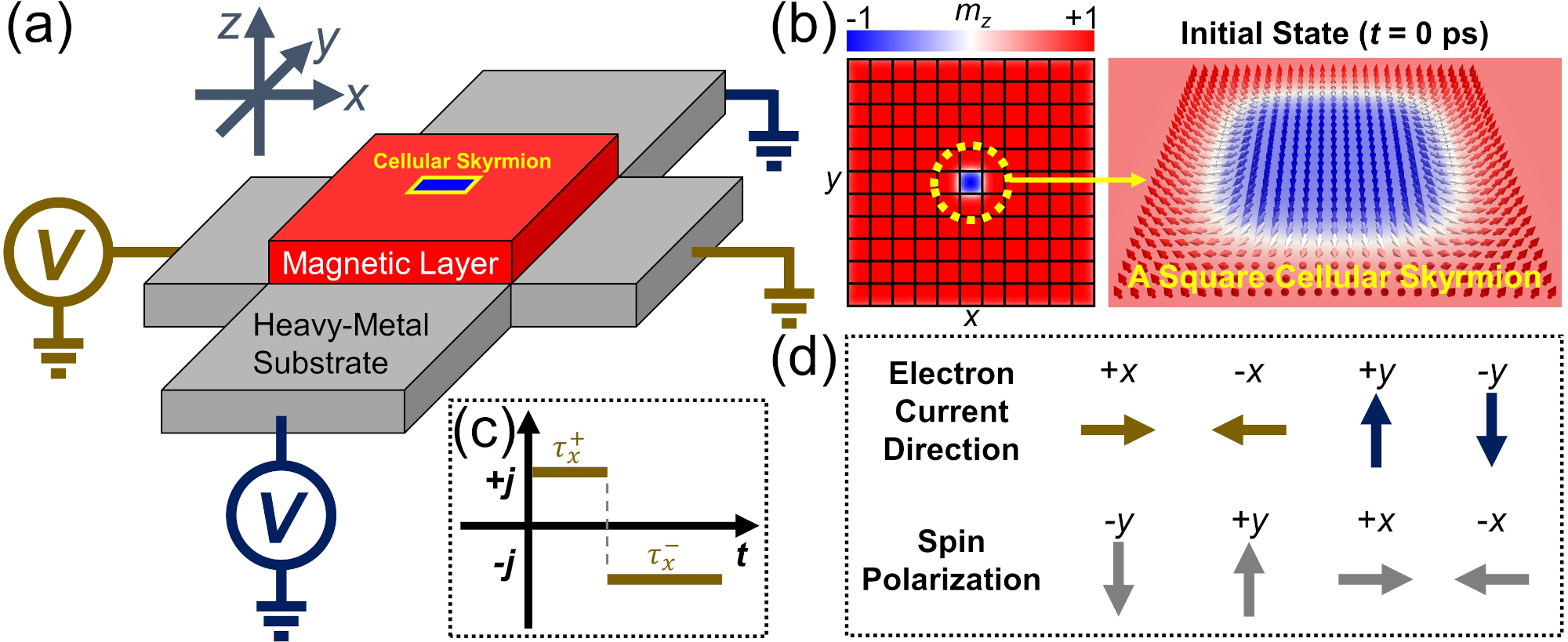}}
\caption{%
(a) Schematic view of the device geometry. A magnetic layer with artificial pinning pattern is attached to a heavy-metal substrate. The device is functioned by currents injected in the $\pm x$ and $\pm y$ directions, which create a spin current propagating into the magnetic layer due to the spin Hall effect.
(b) Top view of the initial state at $t=0$ ps. A square cellular skyrmion is placed and relaxed at the center of the magnetic layer. The color scale represents the out-of-plane magnetization component $m_z$. The black lines indicate the pinning lines with reduced anisotropy.
(c) Schematic of an exemplary programmed pulse profile. A pulse is applied in the $+x$ direction, followed by a pulse applied in the $-x$ direction. The pulse length is indicated by $\tau$, with the subscript and superscript denoting the direction and sign, respectively.
(d) The spin polarization direction $\boldsymbol{p}$ depends on the direction of the electron current $\boldsymbol{\hat{j}}$, which is along $\boldsymbol{\hat{j}}\times\boldsymbol{n}$, where $\boldsymbol{n}$ stands for the surface normal.
}
\label{FIG1}
\end{figure*}

\begin{figure*}[t]
\centerline{\includegraphics[width=0.99\textwidth]{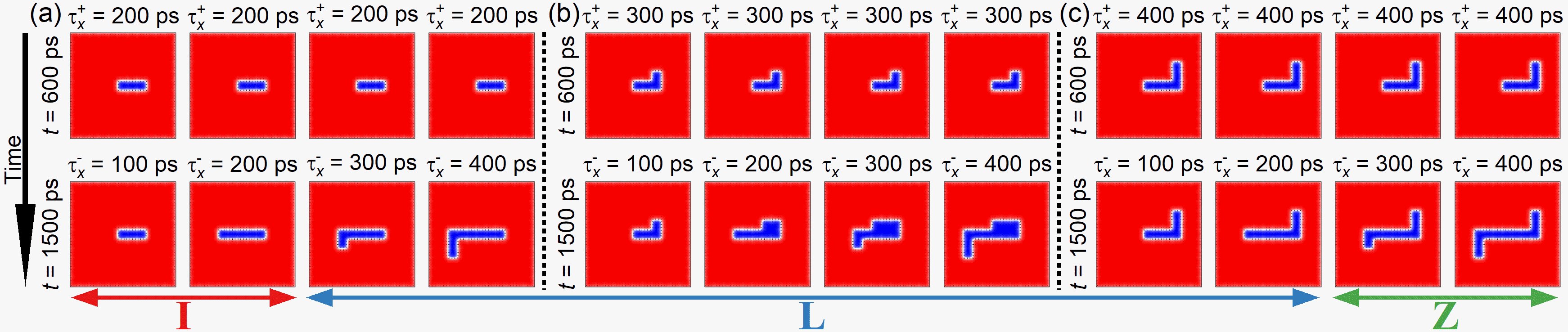}}
\caption{%
Transformation of a square cellular skyrmion induced by parallel current pulses.
(a) Top views of the magnetic layer at selected times (i.e., $t=600$ ps and $t=1500$ ps), showing the relaxed states after the injection of positive and negative pulses. The initial state at $t=0$ ps is given in Fig.~\ref{FIG1}(b). A positive pulse of $\tau_{x}^{+}=200$ ps is applied in the $+x$ direction \magenta{during the time range of $100$ ps $\leq t\leq$ $300$ ps}, followed by a relaxation until $t=600$ ps. Then, a negative pulse of $\tau_{x}^{-}=100$-$400$ ps is applied in the $-x$ direction, followed by a relaxation until $t=1500$ ps. The states at $t=600$ ps and $t=1500$ ps are relaxed states. Here, the current densities of the positive and negative pulses are equal to $j=100$ MA cm$^{-2}$.
(b) A positive pulse of $\tau_{x}^{+}=300$ ps is applied \magenta{during the time range of $100$ ps $\leq t\leq$ $400$ ps}, followed by a relaxation until $t=600$ ps. Then, a negative pulse of $\tau_{x}^{-}=100$-$400$ ps is applied, followed by a relaxation until $t=1500$ ps.
(c) A positive pulse of $\tau_{x}^{+}=400$ ps is applied \magenta{during the time range of $100$ ps $\leq t\leq$ $500$ ps}, followed by a relaxation until $t=600$ ps. Then, a negative pulse of $\tau_{x}^{-}=100$-$400$ ps is applied, followed by a relaxation until $t=1500$ ps.
}
\label{FIG2}
\end{figure*}

\begin{figure}[h]
\centerline{\includegraphics[width=0.50\textwidth]{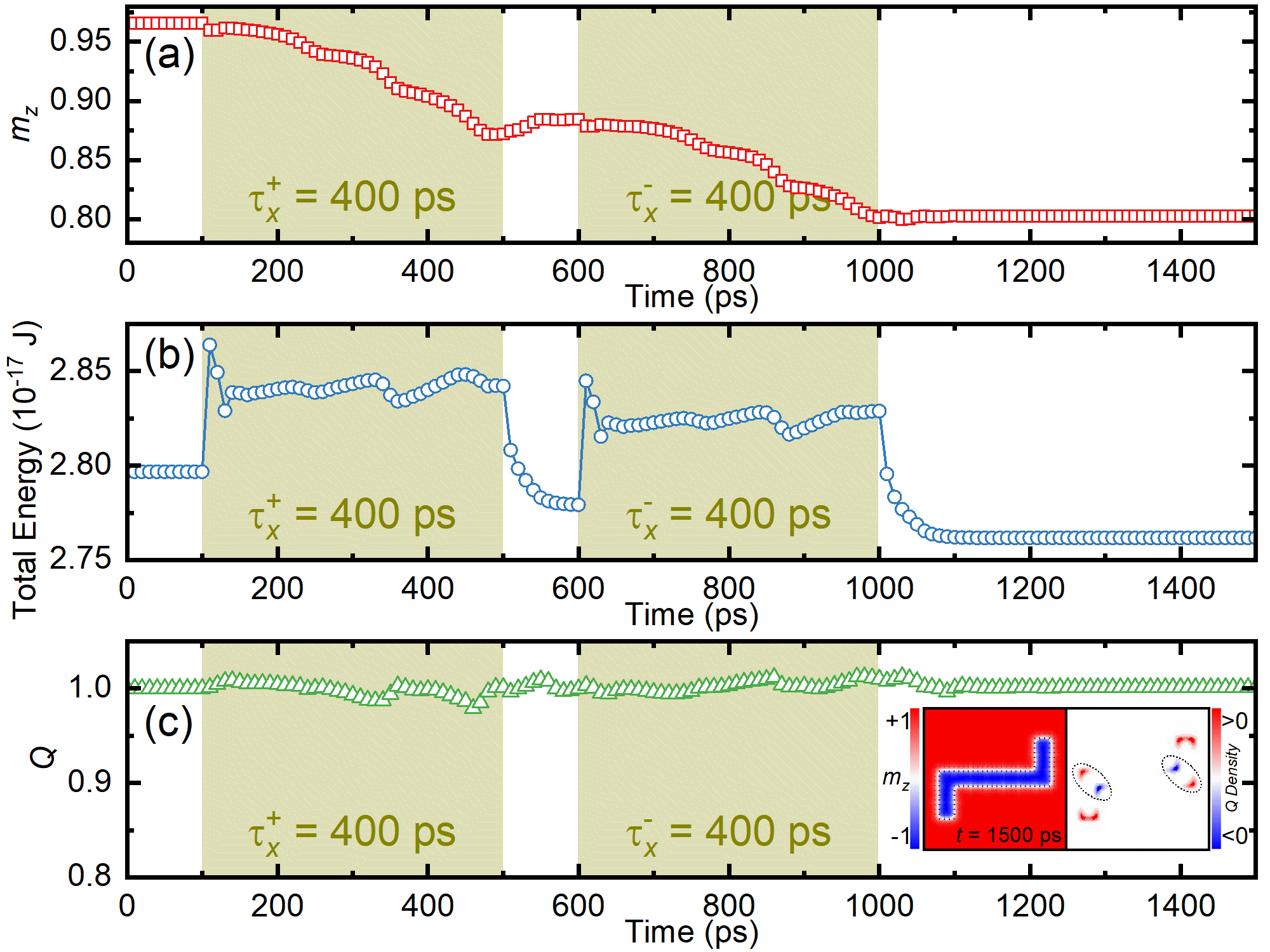}}
\caption{%
(a) Time-dependent reduced out-of-plane magnetization $m_z$ during the current-induced transformation of a square cellular skyrmion to a typical Z-shaped polyomino-like structure, corresponding to the case given in Fig.~\ref{FIG2}(c) with $\tau_{x}^{+}=\tau_{x}^{-}=400$ ps.
(b) Time-dependent total energy of the system.
(c) Time-dependent numerical topological charge $Q$ of the system.
The inset shows the top views of the relaxed Z-shaped polyomino-like structure at $t=1500$ ps and its topological charge density distribution. Two pairs of positive and negative local topological charge densities are indicated by dashed line circles, which could be completely compensated. The system carries a net topological charge $Q=1$, identical to the initial state.
}
\label{FIG3}
\end{figure}

Figure~\ref{FIG1}(a) depicts the device geometry.
We consider a magnetic layer with interface-induced DM interactions and perpendicular magnetic anisotropy.
The magnetic layer size is $378\times 378\times 1$ nm$^{3}$.
We also consider a grid pinning pattern in the magnetic layer. The pinning line width is $4$ nm. The spacing between the edges of two adjacent parallel pinning lines equals $30$ nm. Therefore, the magnetic layer includes $121$ (i.e., $11\times 11$) unit grid cells to host cellular skyrmionic structures.
\emph{The grid pinning pattern can be created by fabricating orthogonal lines with reduced perpendicular magnetic anisotropy~\cite{Ohara_NL2021,Juge_NL2021,Jong_PRB2022}.}

\emph{In experiments, it is possible to locally modify the perpendicular magnetic anisotropy in a magnetic layer to create defect or pinning lines by using the focused helium ion beam irradiation or by fabricating patterned low-dimensional multilayer structures~\cite{Ohara_NL2021,Juge_NL2021,Jong_PRB2022}.}
\emph{A recent experimental work also demonstrated that topographically modifying the substrate could be an effective approach to introduce pinning to the magnetic layer, which was realized by first patterning indentations onto a standard substrate, followed by depositing magnetic layers using magnetron sputtering~\cite{Zhao_SB2024}.}

As shown in Fig.~\ref{FIG1}(b), the initial state at $t=0$ ps is a relaxed square cellular skyrmion placed at the center of the magnetic layer. The cellular skyrmion shows a square shape, which is a result of the attractive interactions between the domain walls and pinning lines. The topological charge carried by the cellular skyrmion is defined as
$Q=-\frac{1}{4\pi}\int\boldsymbol{m}\cdot(\frac{\partial\boldsymbol{m}}{\partial x}\times\frac{\partial\boldsymbol{m}}{\partial y})dxdy=1$,
where $\boldsymbol{m}$ is the reduced magnetization.
We focus on the situation that the grid pinning pattern creates a moderate pinning effect that avoids the translational motion of the cellular skyrmion but favors its deformation or transformation. 

To drive the dynamics of the cellular skyrmion, we assume that the driving electron current is injected into the heavy-metal substrate underneath the magnetic layer. The current can be injected into the heavy metal in the $\pm x$ and $\pm y$ directions [Figs.~\ref{FIG1}(a) and~\ref{FIG1}(c)], which can generate spin currents propagating into the magnetic layer with different polarization directions [Fig.~\ref{FIG1}(d)] due to the spin Hall effect~\cite{Sampaio_NN2013,Tomasello_SREP2014,Zhang_CP2021,Zhang_PRB2022A}.
The spin dynamics in the magnetic layer is therefore governed by the Landau-Lifshitz-Gilbert equation augmented with the damping-like spin-orbit torque~\cite{OOMMF}
\begin{equation}
\begin{split}
\label{eq:LLG-SHE}
\frac{d\boldsymbol{M}}{dt}=&-\gamma_{0}\boldsymbol{M}\times\boldsymbol{H}_{\text{eff}}+\frac{\alpha}{M_{\text{S}}}(\boldsymbol{M}\times\frac{d\boldsymbol{M}}{dt}) \\
&+\frac{u}{M_{\text{S}}}(\boldsymbol{M}\times \boldsymbol{p}\times \boldsymbol{M}),
\end{split}
\end{equation}
where $\boldsymbol{M}$ is the magnetization, $M_{\text{S}}=|\boldsymbol{M}|$ is the saturation magnetization, $t$ is the time, $\gamma_{\text{0}}$ is the absolute value of gyromagnetic ratio, $\alpha$ is the Gilbert damping parameter, and $\boldsymbol{H}_{\text{eff}}=-\mu_{0}^{-1} \partial\varepsilon/\partial \boldsymbol{M}$ is the effective field.
We note that $u=|(\gamma_{0}\hbar)/(\mu_{0}e)|\cdot(j\theta_{\text{SH}})/(2aM_{\text{S}})$ is the spin torque coefficient, $\boldsymbol{p}$ stands for the unit spin polarization direction [Fig.~\ref{FIG1}(d)], $\mu_0$ is the vacuum permeability constant, $\hbar$ is the reduced Planck constant, $e$ is the electron charge, $j$ is the applied current density, and $\theta_{\text{SH}}$ is the spin Hall angle.
The field-like torque is excluded as there is no interfacial Rashba effect in our system~\cite{Wanjun_NPHYS2017,Litzius_NPHYS2017}.
The average energy density $\varepsilon$ includes the ferromagnetic exchange, magnetic anisotropy, demagnetization, and DM interaction energy terms
\begin{equation}
\label{eq:energy-density} 
\begin{split}
\varepsilon=&A\left[\nabla\left(\frac{\boldsymbol{M}}{M_{\text{S}}}\right)\right]^{2}-K\frac{(\boldsymbol{n}\cdot\boldsymbol{M})^{2}}{M_{\text{S}}^{2}}-\frac{\mu_{0}}{2}\boldsymbol{M}\cdot\boldsymbol{H}_{\text{d}} \\
&+\frac{D}{M_{\text{S}}^{2}}\left[M_{z}\left(\boldsymbol{M}\cdot\nabla\right)-\left(\nabla\cdot\boldsymbol{M}\right)M_{z}\right],
\end{split}
\end{equation}
where $A$, $K$, and $D$ are the exchange, anisotropy, and DM interaction energy constants, respectively. $\boldsymbol{H}_{\text{d}}$ is the demagnetization field. $\boldsymbol{n}$ is the unit surface normal vector. $M_z$ is the out-of-plane component of $\boldsymbol{M}$.
The simulation parameters are~\cite{Sampaio_NN2013,Tomasello_SREP2014,Zhang_CP2021}: $\gamma_{0}=2.211\times 10^{5}$ m A$^{-1}$ s$^{-1}$, $\alpha=0.3$, $M_{\text{S}}=580$ kA m$^{-1}$, $A=15$ pJ m$^{-1}$, $K=0.8$ MJ m$^{-3}$, $D=3$ mJ m$^{-2}$, and $\theta_{\text{SH}}=0.2$.
The pinning lines have a reduced anisotropy of $K_{p}=0.2K$ to attract domain walls.
\emph{Note that the value of $K_{p}/K$ may affect the critical current density required for the transformation (\blue{see the supplementary material, Fig.~S1}).}
The simulations are carried out by using the Object Oriented MicroMagnetic Framework (OOMMF)~\cite{OOMMF} with a mesh size of $2$ $\times$ $2$ $\times$ $1$ nm$^3$, which guarantees both accuracy and efficiency.

\begin{figure*}[t]
\centerline{\includegraphics[width=0.99\textwidth]{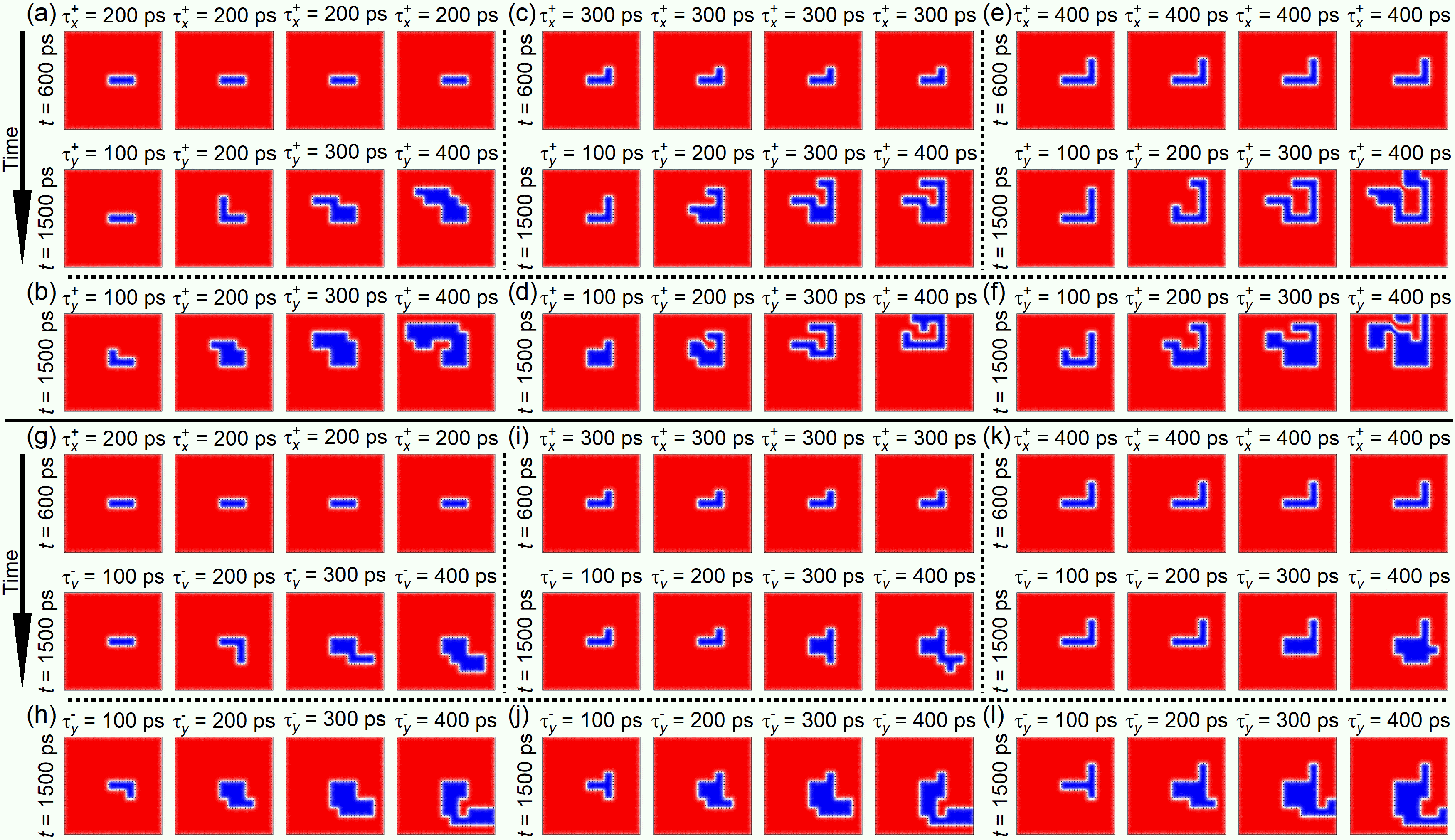}}
\caption{%
Transformation of a square cellular skyrmion induced by orthogonal current pulses.
(a) Top views of the magnetic layer at selected times (i.e., $t=600$ ps and $t=1500$ ps), showing the relaxed states after the injection of two orthogonal pulses. The initial state at $t=0$ ps is given in Fig.~\ref{FIG1}(b). A positive pulse of $\tau_{x}^{+}=200$ ps is applied in the $+x$ direction \magenta{during the time range of $100$ ps $\leq t\leq$ $300$ ps}, followed by a relaxation until $t=600$ ps. Then, a positive pulse of $\tau_{y}^{+}=100$-$400$ ps is applied in the $+y$ direction, followed by a relaxation until $t=1500$ ps. The states at $t=600$ ps and $t=1500$ ps are relaxed states. The current densities of $\tau_{x}^{+}$ and $\tau_{y}^{+}$ are set to $j=100$ MA cm$^{-2}$ and $j=110$ MA cm$^{-2}$, respectively.
(b) The $\tau_{x}^{+}$-$\tau_{y}^{+}$ profile is the same as in (a), but the current density of $\tau_{y}^{+}$ is increased to $j=120$ MA cm$^{-2}$.
(c) A positive pulse of $\tau_{x}^{+}=300$ ps is applied \magenta{during the time range of $100$ ps $\leq t\leq$ $400$ ps}, followed by a relaxation until $t=600$ ps. Then, a positive pulse of $\tau_{y}^{+}=100$-$400$ ps ($j=110$ MA cm$^{-2}$) is applied, followed by a relaxation until $t=1500$ ps.
(d) The $\tau_{x}^{+}$-$\tau_{y}^{+}$ profile is the same as in (c), but the current density of $\tau_{y}^{+}$ is increased to $j=120$ MA cm$^{-2}$.
(e) A positive pulse of $\tau_{x}^{+}=400$ ps is applied \magenta{during the time range of $100$ ps $\leq t\leq$ $500$ ps}, followed by a relaxation until $t=600$ ps. Then, a positive pulse of $\tau_{y}^{+}=100$-$400$ ps ($j=110$ MA cm$^{-2}$) is applied, followed by a relaxation until $t=1500$ ps.
(f) The $\tau_{x}^{+}$-$\tau_{y}^{+}$ profile is the same as in (e), but the current density of $\tau_{y}^{+}$ is increased to $j=120$ MA cm$^{-2}$.
(g) A positive pulse of $\tau_{x}^{+}=200$ ps is applied \magenta{during the time range of $100$ ps $\leq t\leq$ $300$ ps}, followed by a relaxation until $t=600$ ps. Then, a negative pulse of $\tau_{y}^{-}=100$-$400$ ps ($j=110$ MA cm$^{-2}$) is applied, followed by a relaxation until $t=1500$ ps.
(h) The $\tau_{x}^{+}$-$\tau_{y}^{-}$ profile is the same as in (g), but the current density of $\tau_{y}^{-}$ is increased to $j=120$ MA cm$^{-2}$.
(i) A positive pulse of $\tau_{x}^{+}=300$ ps is applied \magenta{during the time range of $100$ ps $\leq t\leq$ $400$ ps}, followed by a relaxation until $t=600$ ps. Then, a negative pulse of $\tau_{y}^{-}=100$-$400$ ps ($j=110$ MA cm$^{-2}$) is applied, followed by a relaxation until $t=1500$ ps.
(j) The $\tau_{x}^{+}$-$\tau_{y}^{-}$ profile is the same as in (i), but the current density of $\tau_{y}^{-}$ is increased to $j=120$ MA cm$^{-2}$.
(k) A positive pulse of $\tau_{x}^{+}=400$ ps is applied \magenta{during the time range of $100$ ps $\leq t\leq$ $500$ ps}, followed by a relaxation until $t=600$ ps. Then, a negative pulse of $\tau_{y}^{-}=100$-$400$ ps ($j=110$ MA cm$^{-2}$) is applied, followed by a relaxation until $t=1500$ ps.
(l) The $\tau_{x}^{+}$-$\tau_{y}^{-}$ profile is the same as in (k), but the current density of $\tau_{y}^{-}$ is increased to $j=120$ MA cm$^{-2}$.
}
\label{FIG4}
\end{figure*}

\begin{figure}[h]
\centerline{\includegraphics[width=0.50\textwidth]{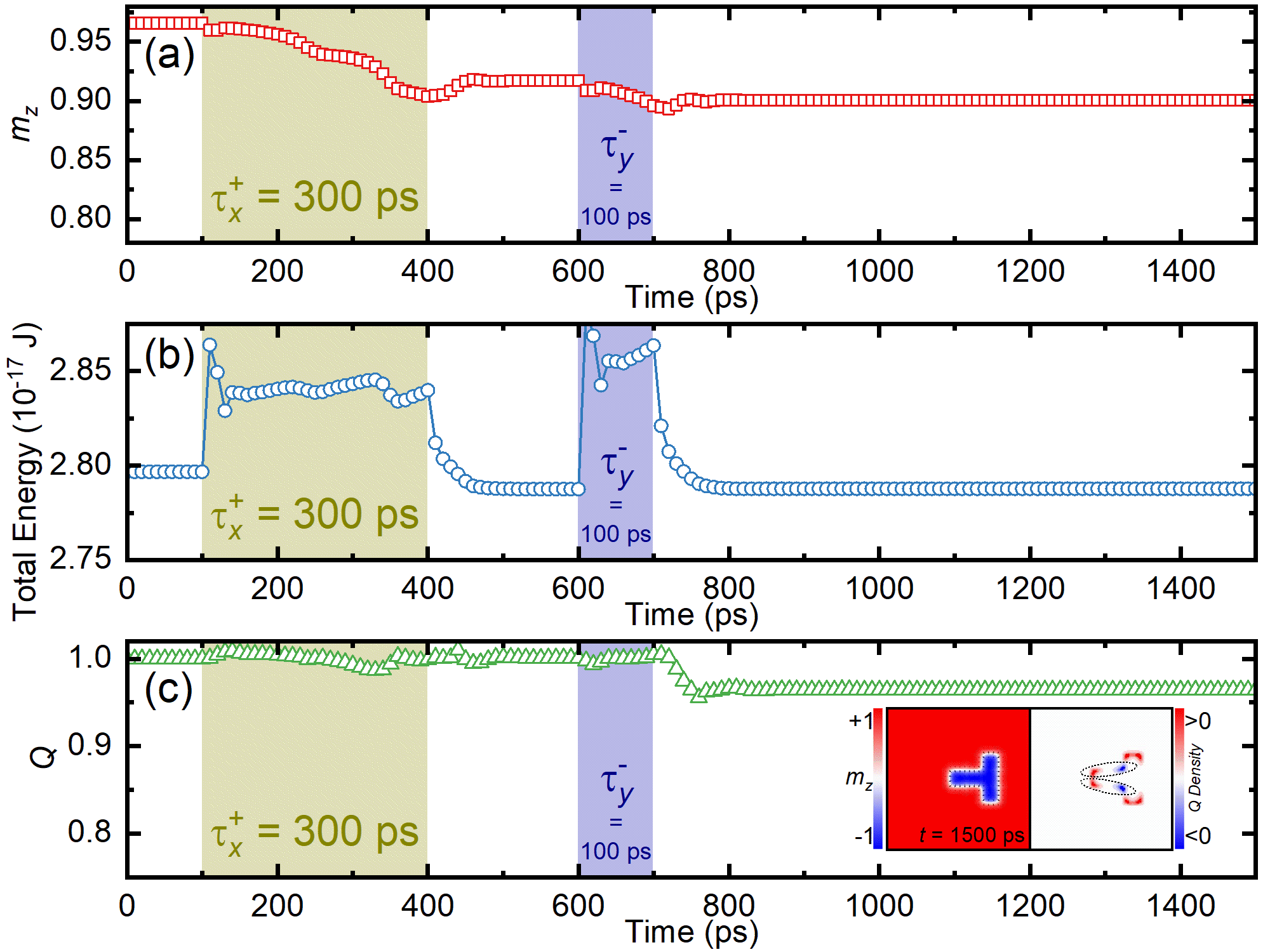}}
\caption{%
(a) Time-dependent reduced out-of-plane magnetization $m_z$ during the current-induced transformation of a square cellular skyrmion to a typical T-shaped polyomino-like structure, corresponding to the case given in Fig.~\ref{FIG4}(j) with $\tau_{x}^{+}=300$ ps and $\tau_{y}^{-}=100$ ps.
(b) Time-dependent total energy of the system.
(c) Time-dependent numerical topological charge $Q$ of the system.
The inset shows the top views of the relaxed T-shaped polyomino-like structure at $t=1500$ ps and its topological charge density distribution. Two pairs of positive and negative local topological charge densities are indicated by dashed line circles, which could not be compensated completely.
}
\label{FIG5}
\end{figure}

We first study the transformation of the square cellular skyrmion induced by a programmed $\tau_{x}^{+}$-$\tau_{x}^{-}$ sequence of parallel pulses [Fig.~\ref{FIG1}(c)].
\magenta{Here, $\tau_{x}^{+}$-$\tau_{x}^{-}$ indicates that a positive pulse with the pulse length $\tau_{x}^{+}$ and a negative pulse with the pulse length $\tau_{x}^{-}$ are applied in a sequential manner.}
\magenta{To be specific, we first apply a positive pulse in the $+x$ direction with $j=100$ MA cm$^{-2}$ and $\tau_{x}^{+}=200$-$400$ ps from $t=100$ ps.}
\magenta{Note that the dash symbol ``-'' indicates that the applied value of the pulse length $\tau_{x}$ is ranged between $200$ and $400$ ps.}
\magenta{The system is relaxed until $t=600$ ps after the application of the first pulse. From $t=600$ ps, we then apply a negative pulse in the $-x$ direction with $j=100$ MA cm$^{-2}$ and $\tau_{x}^{-}=100$-$400$ ps, followed by a relaxation until $t=1500$ ps.}
We find that the system is fully relaxed at $t=600$ ps and $t=1500$ ps after the applications of the first and second pulses, respectively.

As shown in Fig.~\ref{FIG2}(a), the square cellular skyrmion is elongated in the $+x$ direction after the first positive pulse with $\tau_{x}^{+}=200$ ps, forming a rectangle polyomino-like structure. After the second negative pulse with $\tau_{x}^{-}=100$-$200$ ps, the rectangle skyrmion is elongated in the $-x$ direction (\blue{see the supplementary material, Video~1}).
This result suggests that a short pulse can lead to a simple elongation of the cellular skyrmion, and the elongation direction can be controlled by the direction of the current.
However, by applying a second negative pulse with $\tau_{x}^{-}=300$-$400$ ps, the rectangle skyrmion is elongated first in the $-x$ direction and then in the $-y$ direction, leading to the formation of an L-shaped polyomino-like structure (\blue{see the supplementary material, Video~2}).
The elongation in the $-y$ direction is due to the skyrmion Hall effect~\cite{Wanjun_NPHYS2017,Litzius_NPHYS2017}, namely, the Magnus force-induced motion of topological spin structures in the direction perpendicular to the driving current. The skyrmion Hall effect is suppressed by the grid pinning pattern due to the attraction between the domain walls and pinning lines~\cite{Zhang_CP2021}. As a result, only a relatively stronger pulse can induce the elongation of polyomino-like structures toward the transverse direction.
For this reason, a longer single positive pulse with $\tau_{x}^{+}=300$-$400$ ps applied in the $+x$ direction can also lead to the formation of an L-shaped polyomino-like structure [Figs.~\ref{FIG2}(b) and~\ref{FIG2}(c)].

Based on the above facts, we demonstrate the possibility to construct a Z-shaped polyomino-like structure by applying a $\tau_{x}^{+}$-$\tau_{x}^{-}$ sequence of pulses [Fig.~\ref{FIG2}(c)], where we set $\tau_{x}^{+}=\tau_{x}^{-}=400$ ps.
The first pulse leads to an L-shaped structure at $t=600$ ps, and the second pulse transforms the L-shaped structure into a Z-shaped one; \blue{see the supplementary material, Fig.~S2 and Video~3}.
During the current-induced formation of the Z-shaped structure, the out-of-plane magnetization of the system decreases with increasing size of the polyomino-like structure [Fig.~\ref{FIG3}(a)].
The total energy of the system is slightly reduced at $t=1500$ ps compared to the initial-state value at $t=0$ ps, indicating the Z-shaped structure is more stable than the initial square skyrmion [Fig.~\ref{FIG3}(b)].
The topological charge of the system slightly varies during the pulse application but both the final Z-shaped structure and initial square skyrmion carry the same topological charge of $Q=1$, justifying the skyrmionic nature of the Z-shaped structure [Fig.~\ref{FIG3}(c)].
However, the Z-shaped structure has both positive and negative local topological charge densities [Fig.~\ref{FIG3}(c) inset], while a square skyrmion only has positive local topological charge density.

We also explore the transformation of the square cellular skyrmion induced by a programmed $\tau_{x}^{+}$-$\tau_{y}^{\pm}$ sequence of orthogonal pulses.
As shown in Fig.~\ref{FIG4}, we first apply a positive pulse in the $+x$ direction with $j=100$ MA cm$^{-2}$ and $\tau_{x}^{+}=200$-$400$ ps from $t=100$ ps. The system is relaxed until $t=600$ ps after the application of the first pulse.
From $t=600$ ps, we then apply a pulse in the $\pm y$ direction with $j=110$-$120$ MA cm$^{-2}$ and $\tau_{y}^{\pm}=100$-$400$ ps, followed by a relaxation until $t=1500$ ps. We find that the system is fully relaxed at $t=600$ ps and $t=1500$ ps after the applications of the first and second pulses, respectively.

As discussed above, the first pulse with $\tau_{x}^{+}=200$-$400$ ps can create an elongated I-shaped (i.e., rectangle) or L-shaped polyomino-like structure at $t=600$ ps.
By applying the second pulse with $\tau_{y}^{\pm}=100$-$400$ ps in the orthogonal directions, we find that the I-shaped and L-shaped polyomino-like structures can be transformed into \emph{other polyomino-like} or irregular-shaped structures [Figs.~\ref{FIG4}(a)-(l)], depending on the length and magnitude of the second pulse.

Here, we focus on non-trivial results, that is, the transformation of the square skyrmion to polyomino-like structures.
For example, by applying a $\tau_{x}^{+}$-$\tau_{y}^{-}$ pulse sequence, as shown in Fig.~\ref{FIG4}(j), where we set $\tau_{x}^{+}=300$ ps ($j=100$ MA cm$^{-2}$) and $\tau_{y}^{-}=100$ ps ($j=120$ MA cm$^{-2}$), the initial square cellular skyrmion is transformed to a T-shaped polyomino-like structure (\blue{see the supplementary material, Fig.~S3 and Video~4}).
During the transformation, the out-of-plane magnetization decreases with increasing size of the polyomino-like structure [Fig.~\ref{FIG5}(a)].
The total energy of the final state is slightly smaller than that of the initial state [Fig.~\ref{FIG5}(b)], which indicates the T-shaped structure is more stable.
The topological charge of the system slightly varies during the pulse application, and is decreased from $Q=1$ at the initial state to $Q\sim 0.96$ at the final state [Fig.~\ref{FIG5}(c)], indicating the T-shaped structure carries a non-integer topological charge.
In principle, the polyomino-like structure should carry an integer topological charge in a continuum thin film. However, the T-shaped structure shows certain positive and negative local topological charge densities that form two uncompensated pairs [Fig.~\ref{FIG5}(c) inset]. The reason is that the domain walls forming the T-shaped structure are strongly affected by the attractive pinning lines, which may induce and stabilize a local deformation that leads to non-integer topological charges.

\emph{It should also be noted that a reset operation (i.e., the transformation from a polyomino-like structure to a square skyrmion) could be realized by applying an appropriate magnetic field pulse perpendicular to the magnetic layer, which will shrink the polyomino-like structure and lead to a square cellular skyrmion (\blue{see the supplementary material, Fig.~S4 and Video~5}). Besides, a more straightforward way to reset both the shape and position is to fully polarize the magnetic layer and then, create a new square cellular skyrmion at the initial position using a magnetic tunnel junction device placed upon the magnetic layer.}

In conclusion, we have studied the transformation of a square cellular skyrmion on a gird pinning pattern induced by parallel or orthogonal pulses. The transformation is programmable by applying well-designed pulse sequence, and can lead to the formation of polyomino-like structures. The parallel positive and negative pulses could create I-, L-, and Z-shaped polyomino-like structures, while the orthogonal pulses could create L- and T-shaped polyomino-like structures. The transformation depends on the pulse length and magnitude, and may also result in the formation of irregular-shaped structures if the pulse strength is inappropriate. A suitable pulse strength is vital for the transformation of the square skyrmion to other polyomino-like structures.
\emph{Note that other adjustable parameters, such as the pinning line width (\blue{see the supplemental material, Fig.~S5}), may also affect the current-induced transformation, which could be explored systematically in future research.}

\magenta{As an outlook, we point out that the cellular skyrmions and other polyomino-like topological spin structures stabilized on pinning patterns could, in principle, be used for multi-state storage and computing devices, where the information processing depends on the manipulation of not only the position but also the size and shape of the information carriers. In particular, the multi-state spintronic device based on different polyomino-like topological spin structures could be important for building artificial synapses~\cite{Chen_MT2023}, where cellular and pixelated topological spin structures on pinning patterns offer the possibility to realize the state transitions in a step-like manner, mimicking the synaptic potentiation and depression functions~\cite{Song_NE2020}. The polyomino-like topological spin structures may also be used as reconfigurable spin wave waveguides in spintronic devices. It is worth mentioning that a recent experimental work has realized the stabilization and bit-shift motion of skyrmions on square arrays of pinning patterns~\cite{Zhao_SB2024}. It will be important to realize the transformations among different topological spin structures on pinning patterns in experiments.}

\emph{On the other hand, in principle, the transformation to polyomino-like topological structures or solitons are possible in both two-dimensional and three-dimensional systems~\cite{Sutcliffe_2012}. Future research may focus on the creation of more promising and exotic three-dimensional structures with non-trivial topology that are of fundamental interest and of practical importance for advanced spintronic devices.}

\vbox{}

See the supplementary material for \emph{additional simulation results and} videos showing the current-induced transformation of a square cellular skyrmion.

\vbox{}
%
This work was supported by the National Natural Science Foundation of China (Grants No. 12104327, No. 12374123, No. 51571126, No. 51771127, and No. 51772004).
X.Z. and M.M. acknowledge support by CREST, the Japan Science and Technology Agency (Grant No. JPMJCR20T1).
\emph{M.M. also acknowledges support by the Grants-in-Aid for Scientific Research from JSPS KAKENHI (Grants No. JP20H00337, No. JP23H04522, and JP24H02231), and the Waseda University Grant for Special Research Projects (Grant No. 2024C-153).}
Y.Z. acknowledges support by the Shenzhen Fundamental Research Fund (Grant No. JCYJ20210324120213037), the Guangdong Basic and Applied Basic Research Foundation (Grant No. 2021B1515120047), the Shenzhen Peacock Group Plan (Grant No. KQTD20180413181702403), and the 2023 SZSTI Stable Support Scheme.
X.L. acknowledges support by the Grants-in-Aid for Scientific Research from JSPS KAKENHI (Grants No. JP21H01364 and No. JP21K18872).
G.Z. acknowledges support by the Central Government Funds of Guiding Local Scientific and Technological Development for Sichuan Province (Grant No. 2021ZYD0025).

\vbox{}\noindent
\textbf{\sffamily AUTHOR DECLARATIONS}

\noindent
\textbf{\sffamily Conflict of Interest}

The authors have no conflicts to disclose.

\vbox{}\noindent
\textbf{\sffamily Author Contributions}

J.X. and X.Z. contributed equally to this work.
J.X., X.Z., and M.M. conceived the idea.
M.M. and G.Z. coordinated the project.
J.X and X.Z. performed the simulation and analyzed the data.
X.Z. and J.X. drafted the manuscript and revised it with input from Y.Z., X.L., G.Z., and M.M.
All authors discussed the results and reviewed the manuscript.

\vbox{}\noindent
\textbf{\sffamily DATA AVAILABILITY}

The data that support the findings of this study are available from the corresponding authors upon reasonable request.


\vbox{}



\end{document}